# Twin Strouhal numbers in pressure loading of circular cylinder at high Reynolds numbers

Øyvind Mortveit Ellingsen[1,2], Xavier Amandolese[1,3], Olivier Flamand[2] & Pascal Hémon[1*]

[1] LadHyX, CNRS-Ecole polytechnique, IP Paris, France

[2] CSTB, Nantes, France

[3] LMSSC, CNAM, Paris, France

[*] corresponding author: pascal.hemon@ladhyx.polytechnique.fr

## Abstract

We present a wind tunnel experiment on a large smooth circular cylinder mounted between walls. The Reynolds number range of interest is [800 000 – 2 170 000]. This low supercritical regime is often encountered in wind engineering applications, especially the excitation of flexible circular structures by vortex shedding. Main measurements are the unsteady wall pressure distributions around the cylinder by means of synchronized pressure taps with high resolution in time and space. By using the bi-orthogonal decomposition of this set of signals, twin Strouhal numbers are detected which correspond to the second and the fourth terms of the decomposition. These terms are found to produce the unsteady lift on the cylinder with a main contribution of the second one for which the associated Strouhal number is close to the one usually found at subcritical Reynolds numbers.

## 1. Introduction

The circular cylinder is the bluff body which is one of the most studied bodies in aerodynamics along years. The circular shape induces indeed fundamental properties of flow, such as stall and unsteady wake and it has a great relevance in engineering applications. Especially in civil engineering there are numerous cases where circular cylinders of various diameters are submitted to wind and excitation by vortex shedding (Demartino & Ricciardelli 2017).

However the flow regime around this bluff body is extremely dependent on the Reynolds number which combines the effect of the cylinder's diameter $D$ and the mean wind velocity $\overline{U}$ such that:

$$Re = \frac{\overline{U}D}{\nu} \qquad (1)$$

where $\nu$ is the air kinematic viscosity, as studied by (James et al. 1980; Warschauer & Leene 1971). As reported by Zdravkovich (1990) the free stream turbulence and surface roughness are other governing parameters. As a first attempt, the drag force coefficient of a 2D smooth circular cylinder in a low turbulence incoming flow is given in Figure 1 versus the Reynolds number, showing the data provided by the Eurocode (2005) and the experimental data from (Achenbach &

Heinecke 1981). For aerodynamic flows of practical applications with $Re$ greater than 50 000, three kinds of regime can be observed, namely subcritical with $Re \lesssim 200\,000$, critical if $200\,000 \lesssim Re \lesssim 600\,000$ and supercritical when $Re \gtrsim 600\,000$ (Hoerner 1965; Simu & Scanlan 1978). Note that the Reynolds number values that we use to define the limit of the three regimes are taken from the Eurocode (2005). While this standard document is supposed to cover various turbulence intensities of the incoming flow, the drag force coefficient evolution with the Reynolds number matches well the data reported in Achenbach & Heinecke (1981), for which the turbulence intensity of the incident flow was close to 0.45%.

In subcritical regime the boundary layer around the cylinder is laminar prior to its separation and the drag force coefficient $C_D = 1.2$. The alternate vortex shedding is well established and the non dimensional frequency of the shedding is given by the Strouhal number

$$St = \frac{f\,D}{\overline{U}} \qquad (2)$$

in which $f$ is the dimensional frequency. In subcritical regime one observes that $St = 0.19 - 0.20$.

As the Reynolds number increases, the boundary layer becomes progressively turbulent at the separation point and the cylinder is subject to the "drag crisis" which characterizes the critical regime. This regime presents large variation of the drag force coefficient that decreases down to 0.4. The alternate vortex shedding is not well organized.

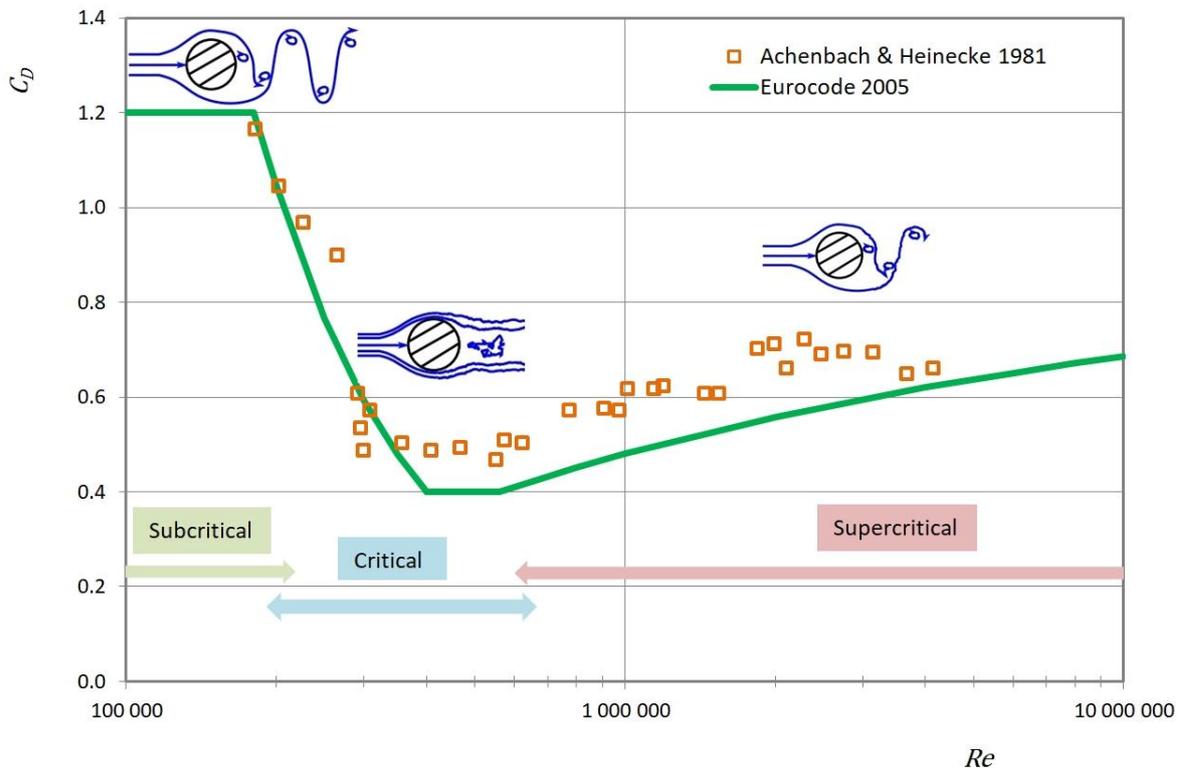

Figure 1: Drag force coefficients from (Achenbach & Heinecke 1981; Eurocode 2005) and definition of Reynolds number regions inspired by (Roshko 1961; Lienhard 1966; Simiu & Scanlan 1978; Blevins 2001).

When the Reynolds number is further increased, reaching the supercritical regime, the boundary layer is fully turbulent prior to separation. This regime is characterized by a smooth monotonic increase of the drag coefficient $C_D$ from 0.4 to 0.65 at $Re = 10^7$. One can also observe a re-organization of the wake with an alternate vortex shedding having a Strouhal number subject to scattering, typically in the range 0.19 – 0.27 as it can be seen in Figure 2 (Rosko 1961; Schewe 1983; Shih et al. 1993; Adachi et al. 1985, Adachi 1997; Zan 2008).

From these data, it turns out obvious that wind tunnel tests have to be made at the right Reynolds number which is encountered in the application. For instance in wind engineering (Lupi et al. 2017; Ellingsen et al. 2021), industrial chimneys have typically a diameter of 2 m and a natural first bending frequency of the order of 1 Hz. Then the critical wind velocity at which the resonance occurs with the alternate vortex shedding is in the range 7.5-10 m/s. The Reynolds number range for this case is then $10^6 – 1.3\,10^6$, which is in the low region of the supercritical regime mentioned above.

But in practice for wind tunnel testing of such structures, scaled models typically of the order of 1/100 are used. This leads to a diameter of the chimney model of 2 cm and requires a wind velocity, in order to comply with the Reynolds number similarity, which is impossible to reach in a subsonic wind tunnel.

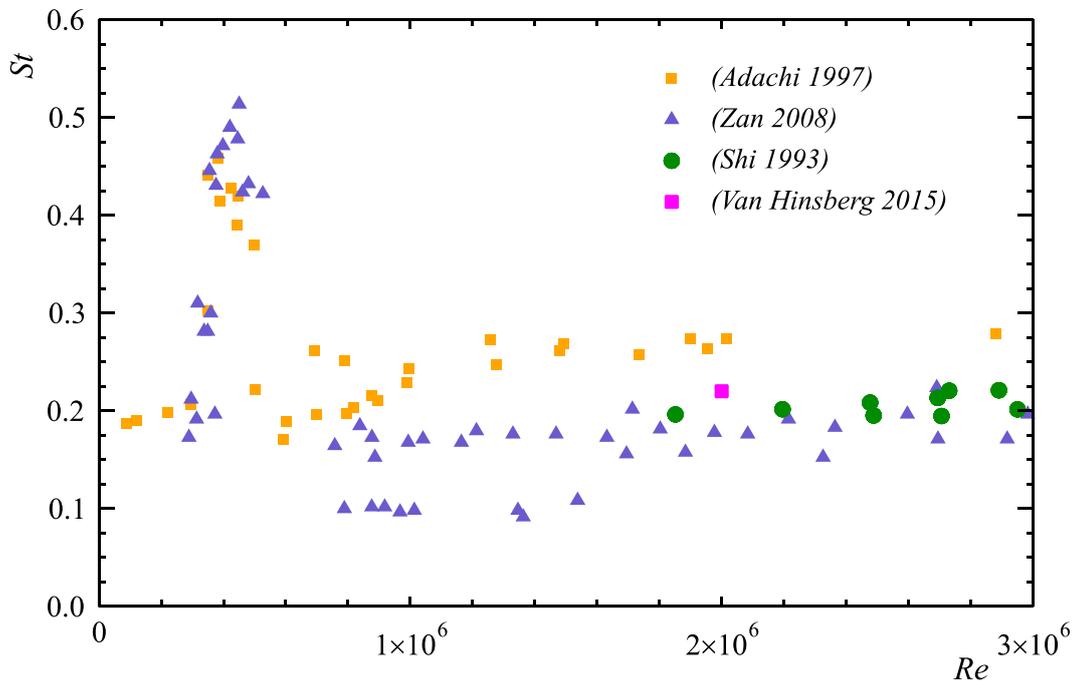

Figure 2. Strouhal number versus Reynolds number from (Adachi 1997; Zan 2008; Shi 1993; Van Hinsberg 2015)

To compensate for this, a number of authors have considered the technique of added roughness on the cylinder model (Achenbach 1971; Szechenyi 1975, Achenbach & Heinecke 1981; Nakamura &

Tomari 1982; Shih et al. 1993; Adachi 1997; van Hinsberg 2015). Rough cylinders are indeed known for shifting the drag crisis at smaller Reynolds numbers, depending on the roughness height. Global parameters such as the drag force coefficient $C_D$, the unsteady lift coefficient (RMS value) $C_\ell'$ and the Strouhal number $St$ are mainly used to calibrate added roughness techniques. But while it is used in wind tunnel testing (Barré & Barnaud 1995), the ability of such techniques to reproduce realistic supercritical flows is still being debated. For example, drag force coefficients with added roughness are always larger than those measured at real supercritical Reynolds numbers.

Recently, van Hinsberg (2015) simultaneously measured the wall pressure distribution and the unsteady forces on a slightly rough circular cylinder up to $Re = 1.2 \times 10^7$. However, the pressure distribution was time-averaged which limits the scope of the study for vortex shedding effects which then still requires investigations. Another interesting experiment was presented in Qiu et al. (2014) but the flow regime was limited to the very beginning of the supercritical regime.

In this paper we present an experimental study which is mainly devoted to the measurement of unsteady wall pressure distributions that form a reference data set for wind tunnel tests on a smooth cylinder submitted to an incident flow of low turbulence intensity up to supercritical Reynolds number. Hence, values of $Re$ up to 2.2 $10^6$ are reached without any artificial techniques. Our range of interest is supercritical with $Re > 800\ 000$. Size of cylinder and flow velocities are fixed so as to fit with those encountered in wind engineering, for instance on industrial chimneys or space launch vehicles when they are subject to vortex shedding excitation.

## 2. Experimental apparatus

The wind tunnel tests were performed in CSTB's climatic wind tunnel in Nantes. The aerodynamic test section, 6x5 m², can reach a maximum wind speed of 70 m/s.

The wind speed is measured by a reference Pitot tube mounted at the entry of the test section, at 0.865 m from the lateral wall and 0.349 m from the roof. A preliminary calibration in the empty test section was performed by mounting another Pitot tube at the cylinder model location to correlate the velocity seen by the model with the one measured by the reference Pitot tube. A maximum uncertainty of 0.5 m/s is estimated for the velocities of interest above 24 m/s.

Homogeneity of the velocity is less than 1 m/s in the test section. The boundary layer on the walls is 0.2 m thick (distance at which 98 % of the reference velocity is reached). The turbulence intensity measured with a cobra probe is almost constant for the high velocities of interest, 1.5 % for the main stream component, 1 % laterally and 1.5 % vertically. In the supercritical regime of interest, i.e. $Re > 800\ 000$, the wake, free shear and boundary layer flows are fully turbulent. Those incident flow turbulence levels should then have negligible effect on the flow regime.

The test model is a smooth circular cylinder with diameter 0.5 m constructed in steel sheet of thichness 3 mm, formed and welded in a specialized factory workshop. It is mounted vertically in the wind tunnel to extend the entire height, as shown in Figure 3. It is clamped at the bottom and supported at its top with a gap with the test section roof less than a millimeter. Note that the clamping setup height is 30 cm (0.6 D) which is slightly larger than the boundary layer thickness.

The cylinder surface roughness for polished steel is of the order of 0.002 mm (Eurocode 2005) which makes a non dimensional rugosity smaller than $10^{-5}$. The cylinder should then be qualified as smooth. The circularity defaults of the cylinder measured outside the wind tunnel was found to be below 1 % of the diameter along its length, leading to a maximum variation of the diameter ±2.5 mm.

The aspect ratio is 10, the same as in (Schewe 1983) and (Van Hinsberg 2015) and larger than (Qiu et al. 2014). It should be noted (see Figure 3) that the clamping setup provides an asymmetry of the cylinder end conditions. Extrinsic three dimensional effects highlighted by Roshko (1993) can not me dismissed then. The cylinder's and wind tunnel's dimensions make the blockage ratio 8.3% and its effect on the Strouhal number should be small (West & Apelt 1982). The setup makes the first natural frequency close to 64 Hz, which represents a minimum reduced frequency of 0.45 at 70 m/s, much larger than the expected Strouhal numbers. Anyway, the associated damping ratio is high, close to 3.64 %, and any residual vibrations should then be of minor impact.

Sixty uniformly spaced pressure taps were drilled at mid-height with an access panel below the sensors in the wake area for easy mounting and troubleshooting (Figure 3). The first pressure tap is placed at the stagnation point ($\theta = 0°$) and the rest spaced out uniformly with a separation of 6°. This spatial resolution is better than the one of previous studies, for instance (Szechenyi 1975) and provides a very good set of data for the unsteady force analysis. Vinyl tubing with length 1.0 m connects the pressure taps to two 32-channel pressure scanners (32HD ESP pressure scanners from Pressure Systems Inc.) with multiplex frequency of 70 kHz. Both pressure scanners were rated up to 7000 Pa and have static errors within ± 0.03 % of full scale. As the tubing can introduce resonance and damping effects which need to be removed, a proprietary transfer function based on the theoretical work of Bergh and Tijdeman (1965) is used in order to recover and correct unsteady pressure distribution until 500 Hz.

Four-hole Cobra probes were used to measure the near wake characteristics, especially the frequency of the vortex shedding. They were mounted at different heights and placed 2 diameters behind the cylinder axis and 0.25 diameter aside, with one probe at the same height as the pressure taps. In the following, only the lateral velocity component is used for the spectral analysis.

During the tests, the wind tunnel speed is kept constant for each measurement point and all the measured signals are recorded during 180 s at the sampling frequency of 400 Hz. As for the spatial resolution of the pressures taps, the record length is high, representing roughly 5000 periods of the vortex shedding at 70 m/s.

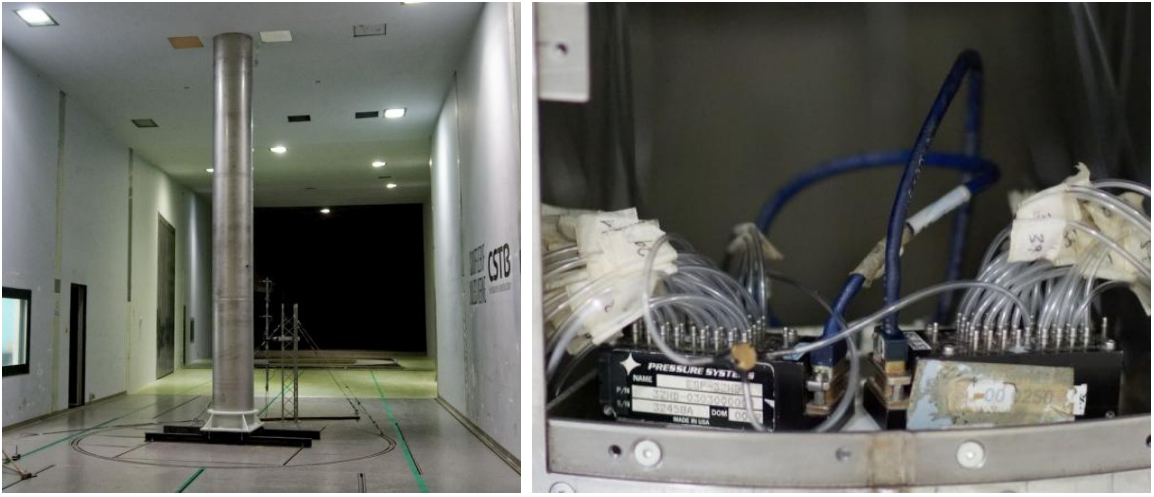

Figure 3: Photo of the large cylinder in the CSTB wind tunnel test section and detailed view of the embedded pressure sensors.

## 3. Wall pressure distribution results

### 3.1. Wall pressure values

The time averaged wall pressure coefficient distribution $\overline{Cp}$ is shown in Figure 4 for three Reynolds number values that span the supercritical regime achieved in the study. The pressure coefficient is defined as

$$Cp(\theta, t) = \frac{P(\theta,t) - P_{ref}}{\frac{1}{2}\rho \overline{U}^2} \tag{3}$$

where $P(\theta, t)$ is the instantaneous measured pressure at the azimuth angle $\theta$. The reference pressure $P_{ref}$ is the mean static pressure in the wind tunnel obtained from the reference Pitot tube and $\rho$ is the air density corrected by atmospheric pressure and air temperature. Maximum uncertainty on pressure coefficient is 1.5 % for this range of Reynolds numbers.

The time averaged distribution presents a classical supercritical shape, with the minimum $Cp_{min} \approx -2.50$ located at $\theta = \pm 80°$. For the highest Reynolds number (2 170 000), the comparison with the results of (Achenbach 1968) at a larger Reynolds number (3 600 000) is relatively good. For these velocities, one can observe an asymmetry of the curves which is, in the present study, attributed to an effect of the cylinder circular section deformation due to the high speed flow (up to 67 m/s). Note that this asymmetry in the mean pressure coefficients distribution provides a sectional lift coefficient close to 0.2 for the highest Reynolds number (2 170 000). It is around 0.05 at low speeds and starts to increase when $Re > 1\ 500\ 000$.

The distribution of the standard deviation of the pressure coefficient $Cp'$ is given in Figure 5 for the same Reynolds numbers. There are two remarkable peaks around $\theta = \pm 110°$ which are the main contributors to the unsteady lift coefficient and linked to the vortex shedding. A deeper analysis of these unsteady signals will be done further in Section 4.

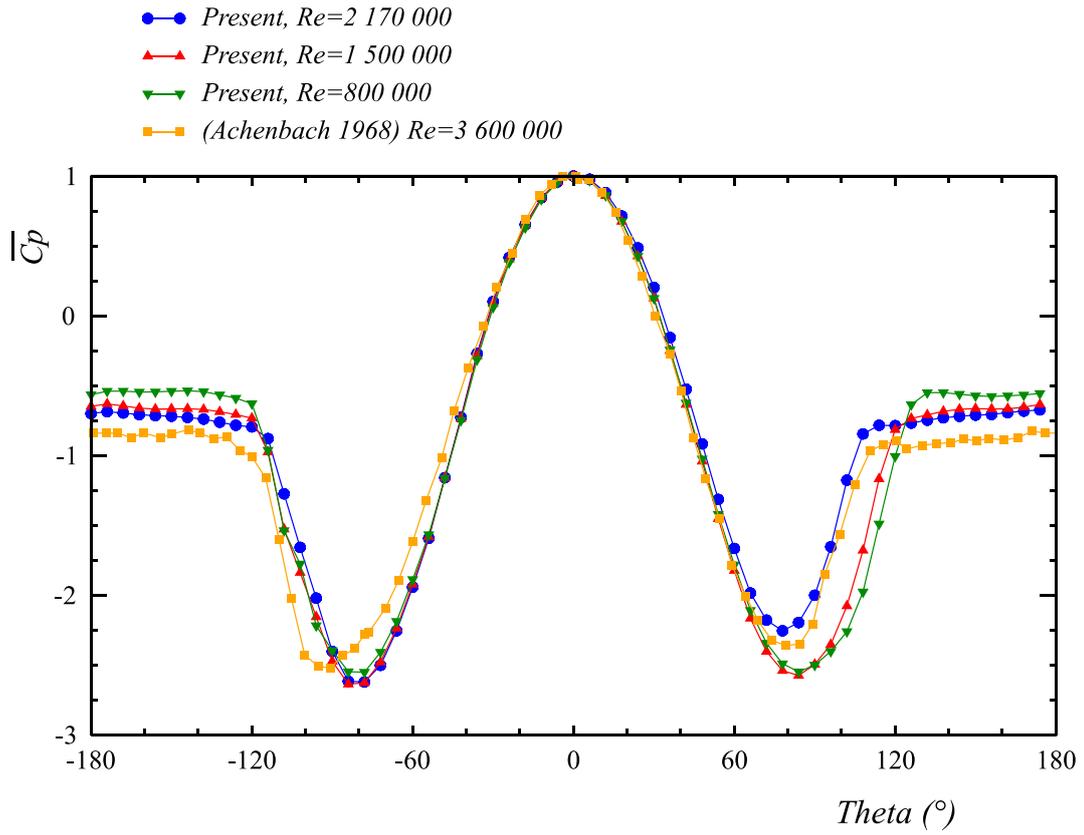

Figure 4. Time averaged wall pressure distribution for different Reynolds numbers

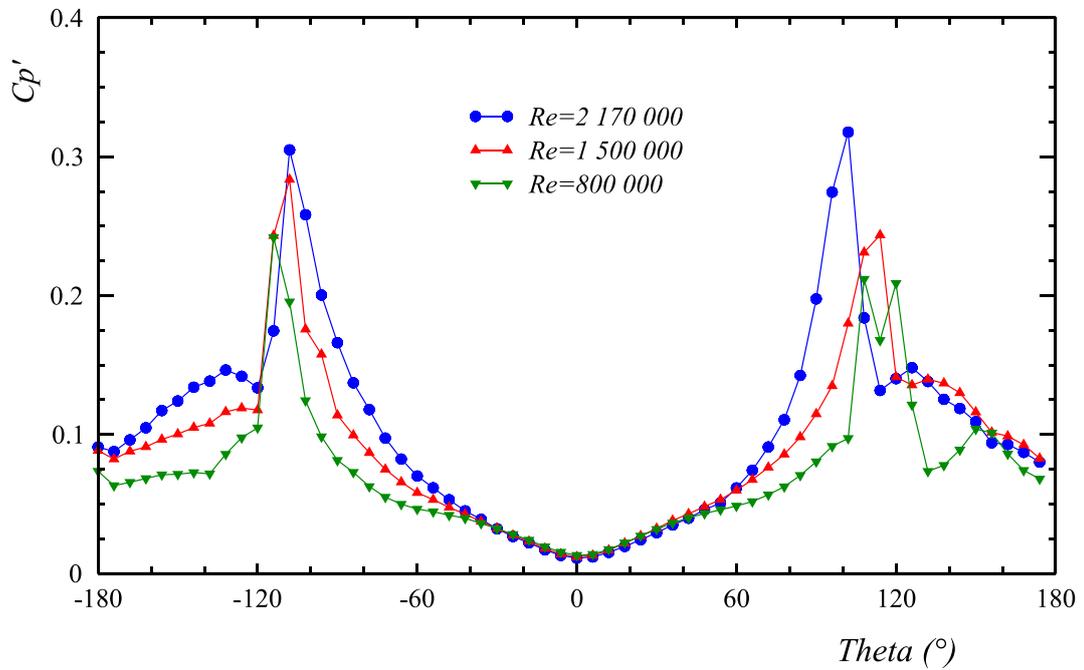

Figure 5. RMS values of wall pressure distribution for different Reynolds numbers

**3.2. Force coefficient values**

The proper integration of the wall pressure signals provides the global force coefficients on the cylinder. Repeatability of some tests led to a maximum difference of 1% for these coefficients. The mean drag coefficient $C_D$ is shown in Figure 6a versus the Reynolds number in the range [800 000 – 2 170 000]. The global trend of the results is quite similar with previous works. Our results are slightly below those of Achenbach and Heinecke (1981). This is probably due to their arrangement with a high blockage ratio (16.7%, twice the current value) which was uncorrected, leading to higher drag coefficient values. However our results match very well the values obtained with the Eurocode formula (2005) and are slightly higher than the one reported in Delany & Sorensen (1953).

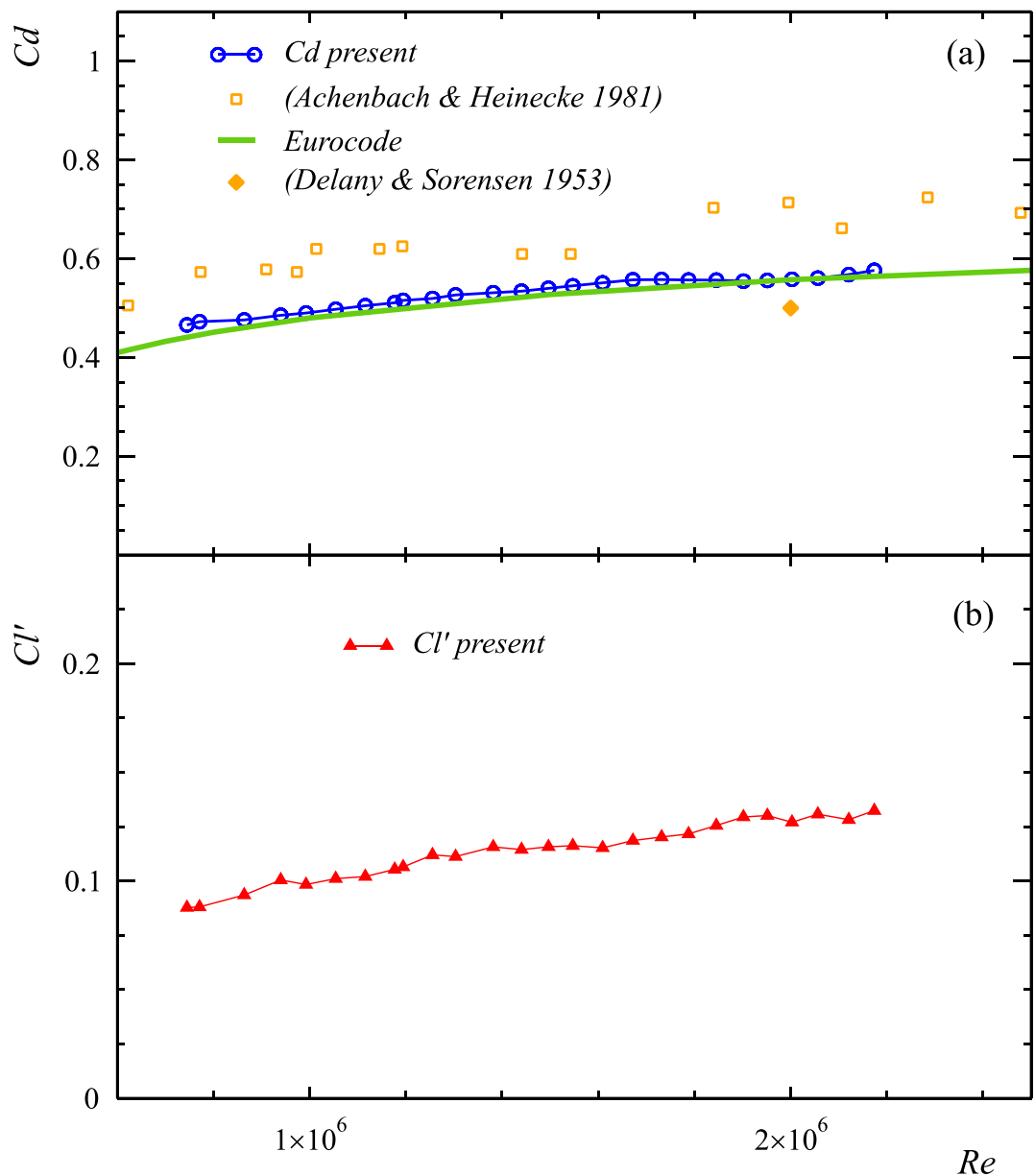

Figure 6. Mean drag coefficient and fluctuating lift coefficient versus Reynolds number

The root-mean square value of the lift coefficient $C_\ell'$ is shown in Figure 6b. At Re=$10^6$ $C_\ell'$ is close to 0.1, that is slightly lower than the value of 0.12 reported by Fung (1960) and greater than Schmidt (1966) who found the range [0.04- 0.095].

## 4. Analysis of the unsteady wall pressure

In this section we use the bi-orthogonal decomposition (BOD) of the wall pressure signals in order to better analyze the alternate vortex shedding which generates the unsteady loading on the cylinder.

### 4.1. The bi-orthogonal decomposition

We recall here the analyzing technique which was first introduced by (Aubry & Lima 1991). The idea of the BOD is to decompose the spatio-temporal signal $Cp(\theta, t)$ in a series of spatial functions $\phi_i(\theta)$ named further as "topos", coupled with a series of temporal functions $\psi_i(t)$ named "chronos". The BOD can be written as

$$Cp(\theta, t) = \sum_{i=1}^{N} \alpha_i \, \phi_i(\theta) \, \psi_i(t) \tag{4}$$

where $\alpha_i$ are the eigenvalues of the spatial or the temporal covariance matrix of the signal $Cp(\theta, t)$. $N$ is the number of terms retained for the decomposition. Chronos and topos are orthogonal between them and normed. Mathematical details can be found in (Aubry & Lima 1991) and practical applications are presented in (Hémon & Santi 2003). It was shown that the eigenvalues $\alpha_i$ are common to chronos and topos and that the series converges rapidly so that $N$ is possibly small compared to the original size $T$ of the problem. $T$ is the smallest value between the number of pressure taps and the number of time records, ie $T = 62$ in the present case. This means that the $\alpha_i$ have a numerical value that decreases rapidly. Their sum

$$A = \sum_{i=1}^{T} \alpha_i \tag{5}$$

represents the total energy in the original signal. Then each couple of chronos and topos have their contribution to the signal which decreases as long as their rank $i$ increases.

It is interesting to analyze the relative contribution of the BOD terms by checking their eigenvalue $\alpha_i$. However, due to the computation technique the first term takes most of the energy of the signal because of the mean value inclusion in the covariance matrix. The complementary reduced energy $1 - \alpha_1/A$ that is related to the unsteady phenomena and can be taken as a basis in order to provide a hierarchy in the following terms of the decomposition. This can be done by calculating their relative contribution $\tilde{\alpha}_i$ to the unsteady energy, which reads:

$$\tilde{\alpha}_i = 100 \frac{\alpha_i/A}{1 - \alpha_1/A} \tag{6}$$

Note that BOD is very similar to proper orthogonal decomposition (POD), except that the mean value of the original signal is kept in the analysis, refer to (Hémon & Santi 2003) for a discussion on that point.

### 4.2. Spatial analysis

The first four topos issued from the BOD of the wall pressure signals at large Reynolds numbers are shown in Figure 7. These first four terms combine 99.8 % of the total energy for the three Reynolds numbers. Further topos of higher rank present noisy shapes and cannot be considered significant in relation to the vortex shedding phenomenon. On the whole, in this low supercritical regime, the Reynolds number does not affect the shape of these first topos. Note that the dissymmetry observed in the pressure distribution is obviously found in these topos.

The evolution of the unsteady energy $\tilde{\alpha}_i$ for the terms 2, 3 and 4 is given in the Figure 8 versus the Reynolds number. As said above, it turns out that the distribution remains almost constant.

The first topos represents the mean value of the wall pressure which produces the mean drag. The topos 2 which the shape is anti-symmetrical is the main contributor to the unsteady lift. The third one is symmetrical and produces an unsteady drag and the topos 4, anti-symmetrical, contributes to the unsteady lift.

The topos 2 is characterized by a sharp peak in the region centred around $\theta = \pm 105°$ with a width of about $\pm 15°$. The fluctuating energy is then mainly located on less than 6 pressure taps, recalling that they are physically spaced by 6°. Although the cylinder surface concerned by this unsteady loading seems small, approximately 2 times 30°, this second term contributes to 32-35 % of the total fluctuating energy.

The topos 3 has a symmetrical shape with two sharp peaks in the same zone as for the topos 2. It corresponds to in-phase pressure fluctuations on both sides of the base region, generating the fluctuating drag force. It contributes to about 20 % of the total fluctuating energy. The fluctuating pressure drag coefficient was found between 0.052 and 0.056 in the range of Reynolds number [800 000 – 2 200 000].

The topos 4 has a wide active region in the rear part with a bump starting at $\theta = \pm 120°$ until the base of the cylinder ($\theta = \pm 180°$). Its anti-symmetrical shape leads to the generation of an unsteady lift component that contributes to 7-8 % of the total fluctuating energy.

Finally, the topos 2 and 4 are those which produce the unsteady lift and store 40-42 % of the fluctuating energy, while the topos 3 produces the unsteady drag with 20 % of the energy.

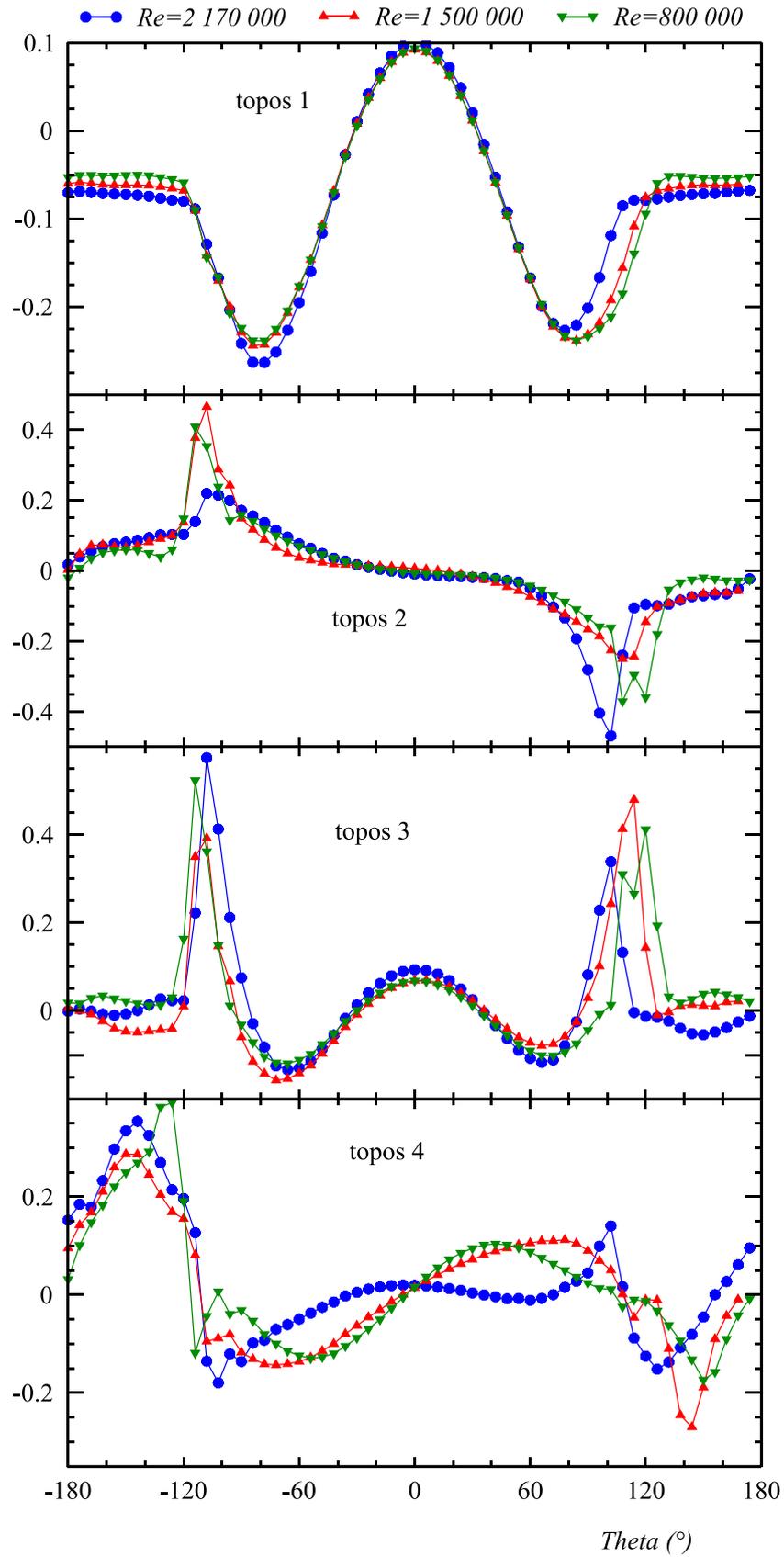

Figure 7. First four topos of the wall pressure at different supercritical Reynolds numbers

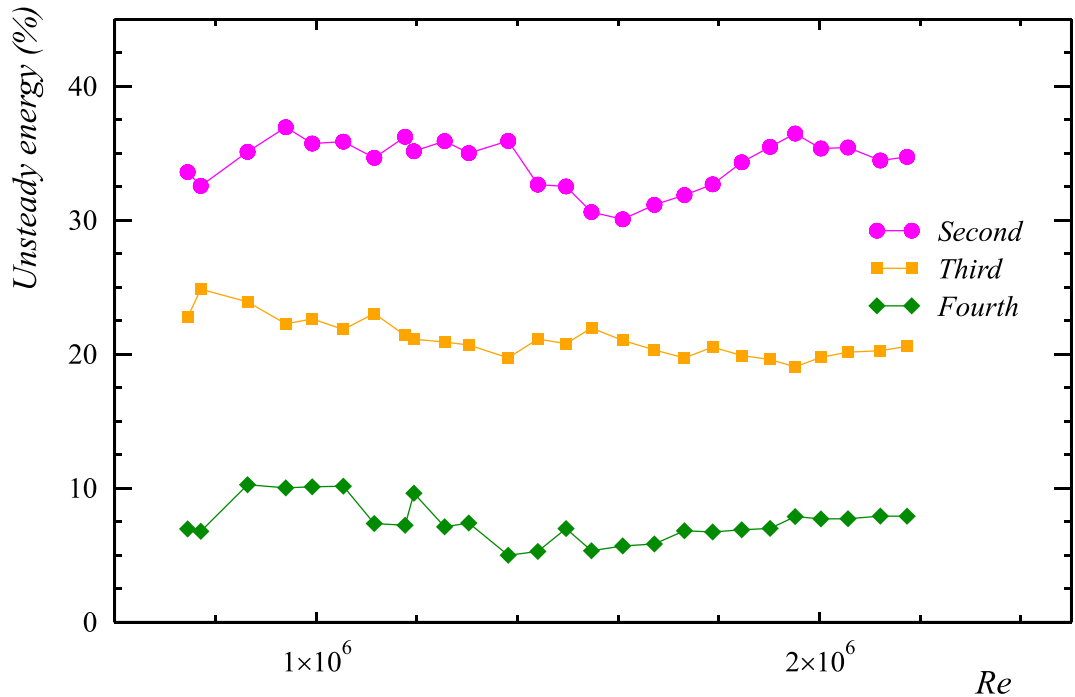

Figure 8. Unsteady energy distribution of the BOD terms 2, 3 & 4 versus Reynolds number

### 4.3. Time analysis

As deduced from the spatial analysis, the chronos associated to the unsteady lift are the second and the fourth one. Their physical length of 180 s sampled at 400 Hz is suitable for an accurate Fourier analysis through the computation of the Power Spectral Density (PSD), using window length of 2048 samples, 50 % overlap and a Hanning tapering function. Final PSD are the average of 70 individual densities. A high pass filter is finally applied in order to cut the energy close to 0 Hz and to concentrate the analysis on the frequencies of vortex shedding.

The time analysis becomes, thus, a frequency analysis which the goal is to obtain the Strouhal number corresponding to the vortex shedding. In Figure 9 the PSD of the cobra probe mounted in the wake and the PSD of the chronos 2 and 4 are presented for the highest Reynolds number of the experiments. It turns out that two peaks are detected in the wake, marked with an arrow, and that these two peaks correspond individually to the peak of each chronos.

Repeating the BOD analysis for all the velocities tested in wind tunnel makes possible to provide the Figure 10 where the two Strouhal numbers are presented versus Reynolds number. The efficiency of the BOD is surprisingly good in separating the two frequencies. Furthermore, the twin Strouhal numbers can be seen as the lower and the higher limit of the Strouhal numbers found by previous authors, (Adachi 1997) and (Zan 2008) especially.

The low Strouhal number which comes from the second term of the decomposition is close to 0.19 – 0.20, the same value which is observed for subcritical Reynolds numbers. It generates the main unsteady lift force that can excite the structure in the case where the critical conditions are reached. Furthermore, we can notice that the low Strouhal number corresponds to Zan's data (2008) from surface pressure measurements, while the current high Strouhal number is more close to Adachi's data (1997) which were measured in the wake.

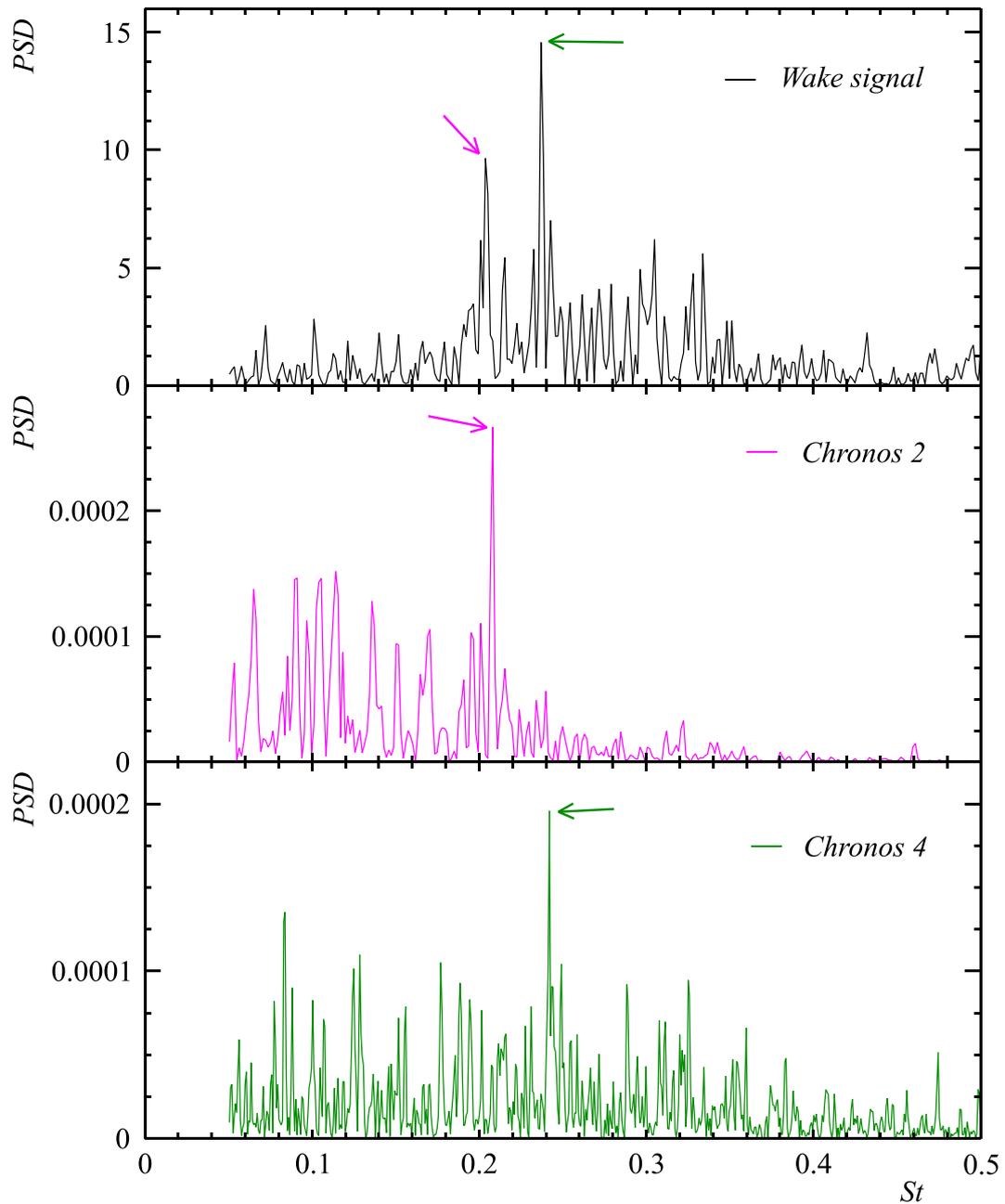

Figure 9. PSD of the wake cobra probe signal and of the chronos 2 & 4 at $Re = 2\,170\,000$

In order to strengthen these findings, we plot in Figure 11 the PSD of the two pressure tap signals at the location of the peak in topos 2, i.e. $\theta = -108°$, and in topos 4, $\theta = -144°$. Although the noise is larger, in comparison with the PSD obtained from the BOD chronos analysis, we clearly find again the individual peaks that correspond to the twin Strouhal numbers.

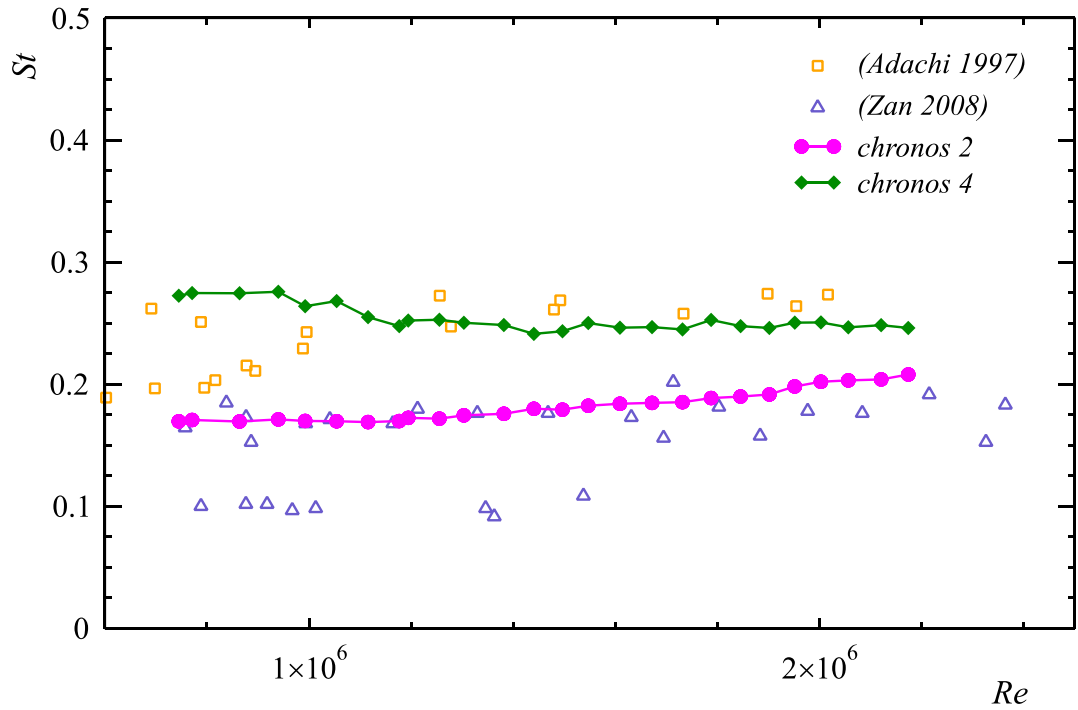

Figure 10. Twin Strouhal numbers versus Reynolds number

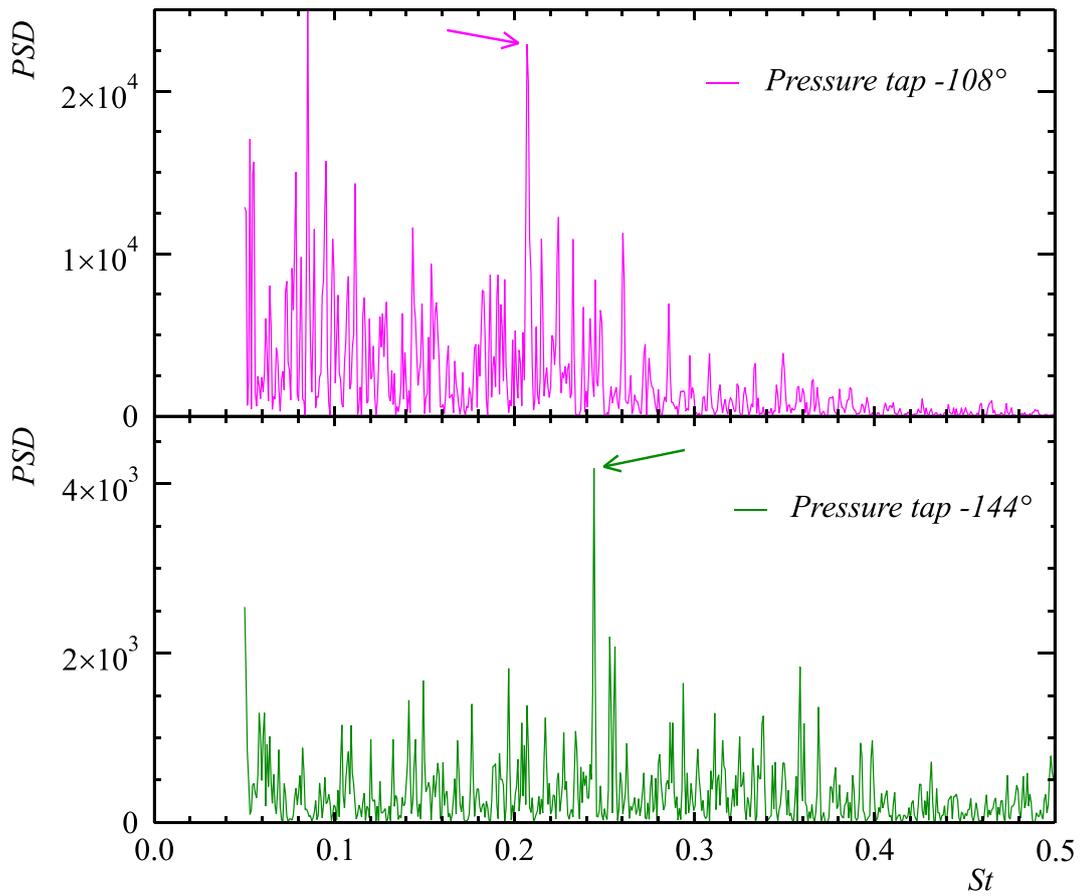

Figure 11. PSD of the pressure tap $\theta = -108°$ and $\theta = -144°$ signals at $Re = 2\,170\,000$

In Figure 10, by looking at the twin Strouhal numbers evolution with the Reynolds number, it seems that the two curves slowly converge and might merge at a higher Reynolds number, reaching a value around 0.22-023 which is sometimes observed by previous authors. Unfortunately the experimental constraints make not possible to reach this regime. There is then the possibility that the twin Strouhal numbers exist only in the currently explored range of Reynolds number, which is however a flow range often encountered in wind engineering applications, especially the vortex shedding excitation of flexible circular structures.

## 5. Conclusion

A large scale experiment was performed on a smooth circular cylinder at supercritical Reynolds number in order to obtain a reference unsteady loading database for this flow regime.

The main findings are summarized in the Table 1. One of the most important results is the detection of twin Strouhal numbers $St_1$ and $St_2$, which is allowed thanks to the high spatial and temporal resolution of the measurements. These two values correspond to the lowest and the highest value found by previous authors. The two frequencies are well separated by means of a bi-orthogonal decomposition of the wall pressure distribution around the cylinder and recovered on the second and the fourth term. The twin Strouhal numbers are also detected by the sensor which was mounted in the near wake, closer to the cylinder than in previous works (Zan 2008; Roshko 1961).

The main unsteady loading due to vortex shedding occurs at the first Strouhal number, taking alone about one third of the unsteady energy of the wall pressure fluctuations. The unsteady pressure loads at the second Strouhal number are lower, taking only 8 % of the unsteady energy. One consequence is that the frequency of the vortex shedding lift force is spread in a much broader band than for subcritical Reynolds numbers.

These new results can be used in further works to validate the ability of the added roughness techniques to simulate supercritical flows and unsteady pressure loading on scaled cylinder models (Ellingsen et al. 2022). End effect and free stream turbulence can also significantly affect the flow regime around smooth circular cylinder. The impact of both those flow-governing factors should then be further clarified for wind engineering application such as industrial chimneys or space launch vehicles.

Table 1: Main characteristics of the cylinder at Re=2 000 000

| | | | |
|---|---|---|---|
| Mean drag $C_D$ | 0.550 | Unsteady lift $C_\ell'$ | 0.127 |
| Strouhal number $St_1$ | 0.20 | Strouhal number $St_2$ | 0.25 |
| $Cp_{min}$ | -2.50 | at $\theta$ | $\pm 80°$ |
| $Cp_{max}'$ | 0.31 | at $\theta$ | $\pm 110°$ |
| Topos 2 $\phi_2(\theta)$ | at $St_1$ | Location of $\phi_2(\theta)_{max}$ | $\pm 110°$ |
| Topos 4 $\phi_4(\theta)$ | at $St_2$ | Location of $\phi_4(\theta)_{max}$ | $\pm 140°$ |

## Acknowledgements

This work is part of a partnership co-funded by Beirens (Poujoulat Group), Centre Scientifique et Technique du Bâtiment (CSTB), Centre National d'Etudes Spatiales (CNES) and LadHyX, CNRS-Ecole polytechnique.